\documentclass[12pt]{article}       

\usepackage[square,numbers]{natbib}

\usepackage[margin=1in]{geometry}
\usepackage{graphicx}
\usepackage{color}
\usepackage[dvipsnames]{xcolor}
\usepackage{authblk}
\usepackage{hyperref}
\hypersetup{breaklinks,colorlinks,allcolors=blue,pdfpagemode=UseNone}
\usepackage{doi}
\usepackage{mathptmx}

\title{Practical Considerations for Data Collection and Management in Mobile Health Micro-randomized Trials}

\author{Nicholas J. Seewald}
\affil{University of Michigan, Department of Statistics}
\author{Shawna N. Smith}
\affil{University of Michigan, Departments of Psychiatry and General Medicine}
\author{Andy Jinseok Lee}
\affil{University of Michigan, School of Information}
\author{Predrag Klasnja}
\affil{University of Michigan, School of Information}
\author{Susan A. Murphy}
\affil{Harvard University, Departments of Statistics and Computer Science}

\begin{document}

\maketitle
\begin{abstract}
There is a growing interest in leveraging the prevalence of mobile technology to improve health by delivering momentary, contextualized interventions to individuals' smartphones. A just-in-time adaptive intervention (JITAI) adjusts to an individual's changing state and/or context to provide the right treatment, at the right time, in the right place. Micro-randomized trials (MRTs) allow for the collection of data which aid in the construction of an optimized JITAI by sequentially randomizing participants to different treatment options at each of many decision points throughout the study. Often, this data is collected passively using a mobile phone. To assess the causal effect of treatment on a near-term outcome, care must be taken when designing the data collection system to ensure it is of appropriately high quality. Here, we make several recommendations for collecting and managing data from an MRT. We provide advice on selecting which features to collect and when, choosing between ``agents'' to implement randomization, identifying sources of missing data, and overcoming other novel challenges. The recommendations are informed by our experience with HeartSteps, an MRT designed to test the effects of an intervention aimed at increasing physical activity in sedentary adults. We also provide a checklist which can be used in designing a data collection system so that scientists can focus more on their questions of interest, and less on cleaning data.
\end{abstract}

\section{Introduction}
\label{sec:intro}
The increasing prevalence of mobile phones and wearable sensors has lead to a great deal of interest in using these technologies to improve health. In particular, ubiquitous computing holds the promise of delivering behavioral interventions that can be tailored to an individual's current context at precisely the right time. A just-in-time adaptive intervention (JITAI) is an emerging mobile health intervention which is intended to provide treatment ``at the right time and in the right place''~\citep{Nahum-Shani2016,Spruijt-Metz2015}. At each of many decision points, a JITAI uses decision rules to determine whether, and if so, which intervention option to deliver to an individual. These decision rules are functions which map data about the individual collected up until the decision point onto an intervention option~\citep{Nahum-Shani2016}. 

The micro-randomized trial (MRT) is an experimental design that provides data for the construction of JITAIs~\citep{Klasnja2015,Liao2016}. Participants in an MRT are sequentially randomized to different treatment options (including no treatment) at each of many decision points at which treatment delivery might be effective.
Often, treatments in this setting are designed to have a nearly real-time impact; as a result, primary analyses for an MRT often focus on the effects of treatment on a ``proximal'' outcome: a near-term, measurable effect of an intervention component through which that component is hypothesized to affect desired distal health endpoints~\citep{Liao2016,Nahum-Shani2016}.
Repeated randomization allows researchers to assess the average causal effects of treatment on proximal outcomes, as well as how these effects change over time and are moderated by participants' context, such as their location, social setting, or mood \citep{Klasnja2015}. The randomizations in an MRT can inform the construction of an effective JITAI by testing intervention components for which there is either insufficient evidence for an effect on a proximal outcome, or for which the dynamics of this effect over time are not well understood~\citep{Klasnja2015}. 

MRTs differ from standard clinical trials in several important ways. First, participants in MRTs may be randomized hundreds or thousands of times at relevant decision points. Second, MRTs are akin to factorial trial designs in that MRTs are designed to provide data that is useful in optimizing/constructing a multifactorial intervention, namely a JITAI.  In an MRT the data analyses focus on examining if the different treatments that might be included in a JITAI have their intended proximal effects. This is very different from a standard clinical trial used to contrast distal health effects of a completely formed intervention (e.g., a JITAI) versus a control. Third, a key goal of MRTs is to inform ``just-in-time'' decision rules---rules for when, where, and for whom a particular intervention component should be delivered---which requires consideration of the specific aspects of context that may affect treatment effectiveness.
In comparison to standard clinical trials, then, MRTs require investigators to collect data at many more time points and, to the extent that contextual moderation effects are of interest, potentially across many more dimensions. Because of the volume of data they need to collect, MRTs frequently make use of automated, passive data collection in order to minimize participant burden. However, MRTs that choose to integrate this passive data collection into the same systems used to deliver the intervention face distinct challenges related to ensuring the collection of high-quality data necessary for evaluating intervention effectiveness and moderation.

Here, we offer practical guidance on data collection and management in an MRT. Specifically, we consider issues related to determining which features to collect, the ``agents'' used to collect those features, differentiating amongst causes of missing data, and novel challenges in preparing MRT data for analysis. Throughout, we provide examples from our experiences with HeartSteps, an MRT designed to optimize a JITAI that used tailored activity suggestions delivered to a participant's smartphone to encourage bouts of physical activity throughout the day. In the Appendix, we present a general checklist which can be used to guide the design of data collection systems in an MRT.

\section{HeartSteps}
\label{sec:heartsteps}

The goal of the HeartSteps project is to develop an effective JITAI for supporting physical activity~\citep{Klasnja2018}. The system deployed in the first HeartSteps MRT consisted of an Android application we developed and the Jawbone Up Move activity tracker that collected minute-level step count data~\citep{Klasnja2015,Smith2017}. The study involved sedentary adults, and one focus of the study was to examine the effect of two different intervention components included in the HeartSteps application: contextually-tailored activity suggestions and activity planning.

Activity suggestions prompted participants to walk or to break sedentary behavior, and they were tailored to the user's current context in order to make them immediately actionable. Suggestions could be provided at five times (decision points) each day and were delivered as notifications to the lock screen of participants' phones. Suggestions remained on the lock screen either until participants interacted with them or until they timed out after 30 minutes. Participants could acknowledge an activity suggestion by giving it a ``thumbs up'' or ``thumbs down'' rating. The five daily decision points were spaced evenly throughout the day, roughly corresponding to morning commute, lunch, mid-afternoon, evening commute, and after dinner. The proximal outcome of interest for the activity suggestions---the outcome that they were intended to directly influence---was the participant's step count in the 30 minutes following the decision point.

\begin{figure}
	\includegraphics[width=\textwidth]{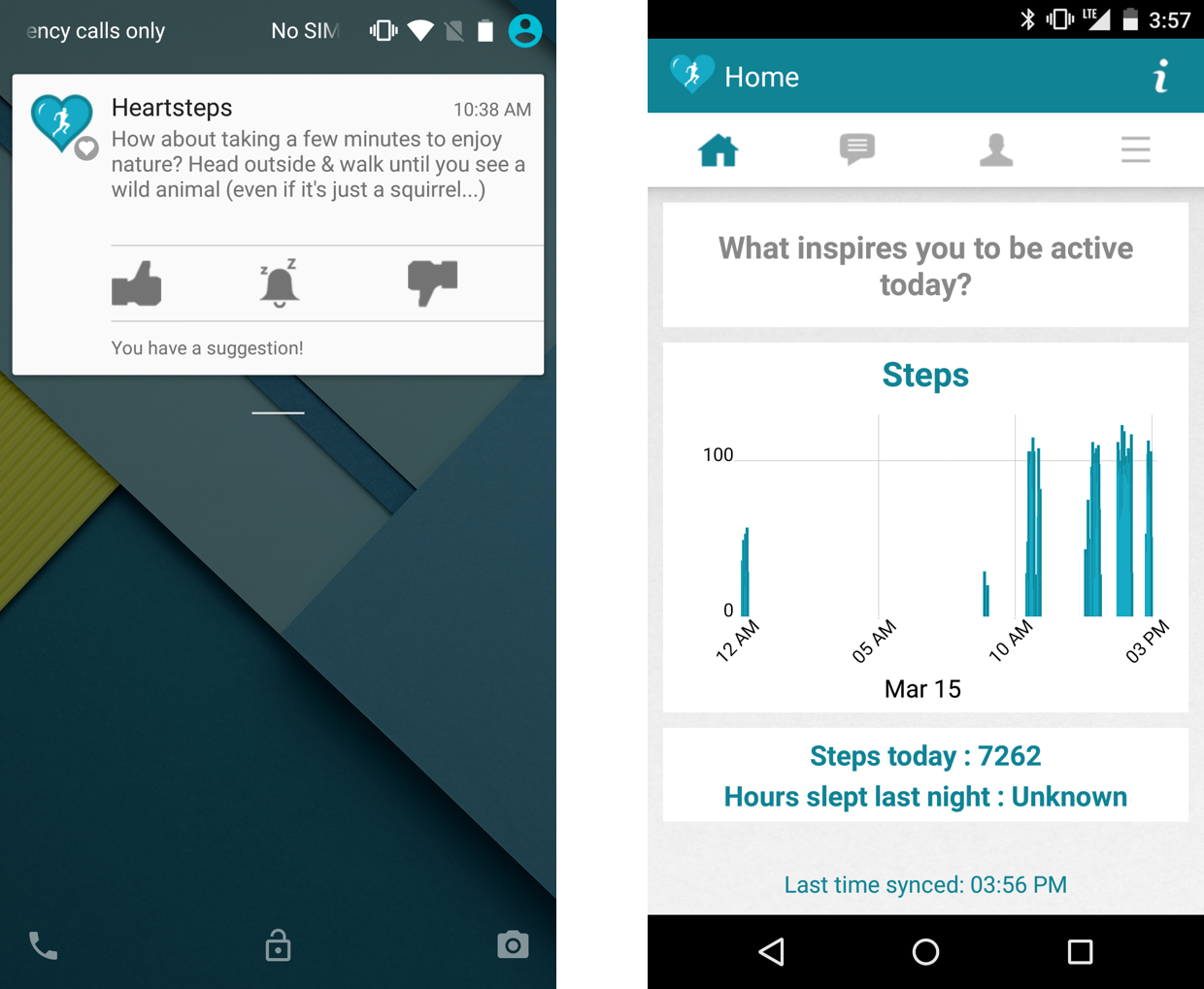}
	\caption{Screenshots from the smartphone app used in HeartSteps. At left, an activity suggestion is delivered to the participant's phone lock screen. The participant can rate the suggestion thumbs-up or -down, or turn off the intervention for up to 12 hours. At right, a screen from the app shows the participant her step counts for the current day.~\protect\citep{Smith2017}}
	\label{fig:hs-screenshots}
\end{figure}

The second treatment assessed in the HeartSteps MRT was activity planning. The planning intervention was designed to help participants specify when, where, and how they would be active on the following day. Planning could be provided each evening as part of the end-of-day questionnaire, which assessed contextual data that could not be captured automatically, such as how stressful and/or hectic the participant's day was or if s/he experienced any illness or travel. The proximal outcome for the evening planning intervention was total step count on the following day. 

The two intervention components above were randomized for each participant during the study as follows: Activity suggestions were randomized to be delivered with probability 0.6 at each of the five decision points each day of the study. Evening planning was randomized to be delivered with probability 0.5 in the evening of each day of the study. Note that the randomization of the planning component was independent of the randomization of activity suggestions. The two components were randomized on different time-scales and had different proximal outcomes. At each decision point for each intervention, we collected data on a number of variables related to treatment delivery and context, resulting in a large, dense dataset. This required careful planning to ensure quality. 

The HeartSteps MRT lasted 42 days and was completed by 37 participants. Each participant experienced at least 210 decision points (5 per day for 42 days) for activity suggestions, and 42 decision points for evening planning. Thus, for 37 participants, there were 7770 possible decision points at which suggestions could be randomized, and a further 1554 possible decision points at which planning could be randomized. The sample size was chosen to achieve 80\% power to detect, for the activity suggestions, an effect size of 0.1, assuming a two-sided type-I error rate of 0.05 and 70\% ``availability'' (see Section \ref{sec:unavail-missingness}) throughout the study~\citep{Klasnja2018}. Under these conditions, the minimum-required sample size is 32 participants~\citep{Liao2016,Seewald2016}. Note that the sample size may appear to be quite small; this is because the primary analyses for MRTs concern main effects of the intervention components in which the primary hypothesis test statistic uses both within- as well as between-person contrasts in proximal outcomes; these types of test statistics are made possible by the within-person randomizations~\citep{Klasnja2015}.

Note that  at each decision point in the HeartSteps MRT treatment assignments were made independently of the individual's past proximal outcomes, past treatment assignment, and context (conditional on them being ``available'' at that decision point; see Section~\ref{sec:unavail-missingness}). This, however, does not impede the study's ability to provide data which can aid in the development of a JITAI. The primary aim of the initial HeartSteps MRT was to examine the average causal effect of sending an activity suggestion (vs. not sending) on the participant's step count in the 30 minutes following the decision point. Secondary aims were to examine how this effect changed over time and/or was moderated by context, as well as to evaluate the time-varying causal effect of evening planning on the next day's step count \citep{Klasnja2018}. Addressing these aims allowed us to determine (1) whether to include each micro-randomized intervention component in the JITAI being developed, and (2) the circumstances under which the intervention component is most effective. For instance, we discovered that suggestions encouraging the participant to walk significantly increased 30-minute post-suggestion step count, whereas the effect for suggestions designed to break sedentary behavior had a smaller, positive, non-significant effect. Furthermore, the effect of the walking suggestions could not be detected by day 29. Subsequent versions of the intervention could be modified to reduce habituation to activity suggestions, which may extend their usefulness~\citep{Klasnja2018}.

\section{Feature Collection}
\label{sec:features}
Passive sensors and mobile phones offer investigators a wide variety of features that can be used to develop a high-quality JITAI. However, restraint must be exercised in choosing which data to collect.
This choice is an important aspect of designing an MRT and, as with any clinical trial, it should be primarily driven by the scientific question(s) of interest. However, the science motivating MRTs is typically concerned with proximal effects of treatments which are often delivered by an application (``app'') on a participant's smartphone in order to achieve timeliness and contextualization. The phone typically also collects the relevant data of interest, both from the treatment and from other sources of (often passively-collected) data. Thus, with MRTs, the choice of which features to collect is inherently linked to both the science and the intervention (app) development process. Scientists should work closely with app developers to ensure that appropriate data is collected and that the system functions reliably in order to ensure the data is high-quality~\citep{Price2014,Smith2017}.

\subsection{Proximal outcome collection}
\label{sec:proximal-outcome}
Particular attention should be given to collecting features that are instrumental in constructing or assessing the primary proximal outcome. As in any clinical trial, the primary outcomes---here, the primary \textit{proximal} outcomes---are specified \textit{a priori} and care should be taken to ensure their proper collection.  For example, when this outcome is measured passively, such as with a sensor or wearable fitness tracker, the scientist must be aware of limitations of both the device itself and of the application programming interfaces (APIs) provided by the manufacturer to access participant data.

\begin{figure}
	\includegraphics[width=\textwidth]{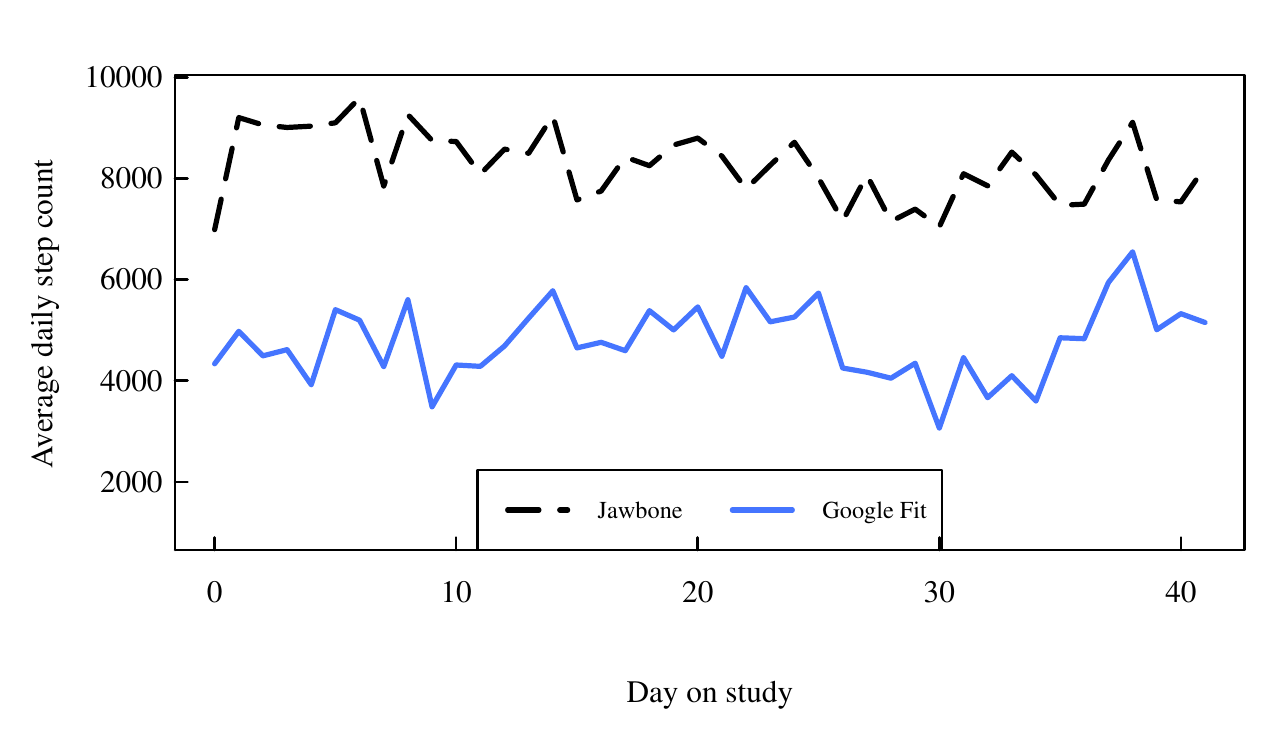}
	\caption{Average daily step counts of HeartSteps participants as measured by Jawbone (dashed black) and Google Fit (solid blue). Google Fit recorded consistently lower step counts than did the Jawbone wearable, but the two data streams follow approximately similar patterns.}
	\label{fig:hs-step-counts}
\end{figure}

Scientists may also consider incorporating redundancy into collection of the primary proximal outcome to ensure robust results. Recall from Section \ref{sec:heartsteps} that HeartSteps measured its proximal outcome of step count primarily using a Jawbone Up Move fitness tracker. Step count data was accessible via a Jawbone-provided API at several time-scales, including minute- and daily-level. We also opted to measure participant step count through the Google Fit app installed on participants' phones. However, Google Fit step count data was less granular than Jawbone's, as the app aggregated step counts across full periods of activity (e.g., from when an individual started walking to when the individual stopped), rather than every minute. Furthermore, Google Fit uses the phone's accelerometer to track step count, which may be less reliable than a wearable sensor (e.g., some individuals may carry their phones in a bag, versus in a pants pocket, which could yield systematically different step counts). Figure~\ref{fig:hs-step-counts} illustrates the difference in step counts as measured by different sensors in HeartSteps.

Nonetheless, having a second measure of step count allowed for greater confidence in the outcome measure and results by enabling certain sensitivity analyses. For one, Google Fit provided a backup source of data when Jawbone data was missing (e.g., when participants forgot to wear their device). In analysis of data from HeartSteps, we performed sensitivity analyses in which missing Jawbone step count data was singly imputed using data from Google Fit, when available~\citep{Klasnja2018}. 
Additionally, collecting another measure of step count provided confidence that our analyses addressed the effect of treatment on physical activity more broadly, rather than simply step count as measured by a particular device. Indeed, there may be notable differences in data provided by different sensors. Step count is particularly susceptible to this and Jawbone accuracy could have been affected by such things as where the tracker is worn (wrist vs. waist) or participant age~\citep{Kumar2013a,Modave2017}. Data analyses, then, might fit models to both the primary outcome and its proxy(ies) to ensure the effects of the intervention are not artifacts of sensor choice. For more on the use of Google Fit data in HeartSteps analyses, see Section \ref{sec:human}.

\subsection{Treatment Delivery}
\label{sec:txt-delivery}
In order to address questions about the effect of a treatment  on its associated proximal outcome, data must be collected on whether, when, and how the intervention was randomized and delivered. The exact information required will vary with the analysis method used, but at minimum, at each decision point the system should store the probability with which the treatment was randomized, the result and time of the randomization, and when and whether the intervention was delivered. Storing randomization probabilities is particularly important: the weighted and centered least-squares approach developed by Boruvka et al. to analyze MRT data centers the treatment indicator with the randomization probability~\citep{Boruvka2017}. While randomization probabilities remained constant in HeartSteps, many MRTs may allow them to change over the study: for example, when interventions are triggered by dynamic behavior, such as sedentary periods that are highly variable both within and between people (see, e.g., Dempsey et al. \citep{Dempsey2017}). Finally, to enable robust causal inference, the same information needs to be recorded at each decision point regardless of whether the treatment is provided. Further discussion of record-keeping in regard to treatment delivery can be found in Section~\ref{sec:agents}. 

To the extent that adherence to treatment is of interest, collection of data related to whether or when the treatment was accessed may also be important. In HeartSteps, for example, while we were able to determine if a participant was randomized to receive an activity suggestion, and whether they rated it, we were not able to directly determine whether they saw or read the activity suggestion (without rating it), nor were we able to sense whether the participant followed the specific suggestion s/he received (although we did observe subsequent step count, as discussed above).

\subsection{Context and Moderators}
\label{sec:context}
A key motivation for building and optimizing JITAIs is the ability to tailor both the content and the delivery of treatment to individuals and their immediate contexts ~\citep{Kumar2013,Nahum-Shani2016}. Tailoring intervention content has been shown to be effective at improving health behavior change, so JITAIs often seek to provide highly context-specific treatments~\citep{Kreuter2000,Noar2011}. For example, the HeartSteps activity suggestions were intended to be immediately actionable, and thus the content of the suggestions was tailored to reflect the participant's current location, time of day, day of the week, and/or weather at each decision point.

Scientists must ensure that apps used in MRTs are able to collect all relevant contextual data in a timely manner. Further, scientists must have protocols in place for dealing with potential lag times in this collection to ensure that the tailored intervention is not using ``stale'' or inappropriate contextual information. For example, in HeartSteps, we did not want to send an activity suggestion that mentioned poor weather when the sun was shining. Anecdotally, this can be jarring to participants, who expect the intervention to be appropriately contextualized~\citep{Smith2017}. Risks of mistailored content can be mitigated by collecting contextual data as close to each decision point as possible, and by having protocols for dealing with stale or absent contextual information (e.g., delivering a generic form of the intervention). 

Contextual data is also useful for discovering the best times to provide treatment by investigating possible moderation effects---potentially time-varying contextual factors that strengthen or weaken intervention effects~\citep{Boruvka2017,Klasnja2015}. For example, with HeartSteps we might hypothesize that providing an evening planning intervention is more effective on weekdays than weekends, or that the effect of activity suggestions declines with time-on-study as participants habituate to the suggestions. Other potential moderators might include participant burden (which could be assessed by the number of treatments provided in some recent window), stress, weather, time of day, or location. 
These moderating effects can easily be assessed using the method developed by Boruvka et al. by including in the ``treatment effects model'' (their equation (5)) an interaction between the moderator of interest and a centered treatment indicator~\citep{Boruvka2017}.

Determining \textit{a priori} which moderator effects are of interest is crucial, so that appropriate data can be collected over the course of the study. Examining time-varying treatment effect moderation requires that data on any and all potential moderators is available at all decision points. For example, to investigate whether stress moderates the impact of an intervention, analysts must have access to a relevant, time-varying measure of stress for every randomization. While this does not necessarily mean that stress should be assessed at every decision point (it could be measured, for instance, every day), the measure used should be scientifically sensible. Note that this also introduces questions regarding the time scales on which variables of interest are available; these are addressed in Section~\ref{sec:time}. 
Investigators interested in a large number of potential moderators should try to pre-specify the relative ``priority'' of each such variable. Moderators analyses may require fitting many models; having an \textit{a priori} hypothesized level of importance for each variable can improve confidence in the results. 
 
\subsection{Participant Privacy}
\label{sec:privacy}
Choosing which features to collect in an MRT often involves a trade-off between the investigator's scientific interests and the participants' privacy. Data collection through passive sensing can hide potential privacy risks from participants~\citep{Raij2011}. Proper anonymization of MRT data is also important, as certain features collected via a smartphone could be used to identify participants. GPS coordinate data, for instance, could be used to infer a participant's home address, even with some form of masking~\citep{Seidl2015}. While the need to exercise restraint in collecting participant data is a challenge for clinical research in general~\citep{Saczynski2013} and not specific to MRTs, privacy risks in mobile health are often greater due to the diversity and temporal density of the collected information. These risks have been much discussed~\citep{Kotz2011,Martinez-Perez2015,Raij2011}, and a number of approaches have been developed for increasing participant privacy, including methods for anonymizing geographic data~\citep{Cassa2008,Seidl2015}. Here, we emphasize mitigation of privacy risks through thoughtful collection of contextual features in an MRT.

Over-collection or improper storage of contextual variables may exacerbate privacy concerns in participants. As discussed above, location data can enable JITAIs to deliver treatments that are highly tailored to a participant's current physical context. However, highly-specific location information is likely of less utility in time-varying moderators analyses. In HeartSteps, we chose to categorize participant location as ``home'', ``work'', or ``other'' to assess effect moderation, ensuring that no specific location information was stored in the long-term. If other contextual moderators that depend on precise location, like weather, are collected accurately and consistently at each decision point, it may be unnecessary to store GPS coordinates or other fine-grained data. We recommend that study designers carefully consider both scientific justifications and participants' privacy concerns when designing MRT data collection systems.

Privacy should also be considered when storing data. Mobile health data might include protected health information (PHI), which may be subject to storage and safety requirements in the United States under the Health Insurance Portability and Accountability Act (HIPAA). If participants' data is sent from their phones to a central server for later analysis, those servers may need to be HIPAA-compliant, and access to the data and PHI should be tightly controlled. However, information that is collected and used solely for contextualizing the intervention may not need to be sent to a server, and could instead stay on the participant's phone, minimizing risk.

\section{Choosing the ``Randomization Agent''}
\label{sec:agents}
Randomization is a key part of MRTs; therefore, choosing \textit{how} to randomize is critical. The technical aspects of this issue are discussed by Smith and colleagues~\citep{Smith2017}; here, we focus on the choice between randomizing treatment assignments on the participant's phone, or on a central server which is also used for data collection. This decision has consequences for both data management and the participant's experience with the intervention. In particular, phone-side randomization can react faster to changing context and provide more precise timing of treatment, but it is also susceptible to a wide variety of technical issues which may interfere with data integrity and/or structure. Conversely, server-side decisions can facilitate a cleaner data structure, but at the expense of speed and precision in timing.

Consider the HeartSteps MRT: in the version of the app we tested in our first study, the decision to push an activity suggestion (and, subsequently, suggestion delivery) was made on the phone. Because the decisions were made ``locally'', an internet connection was not required at the time of the decision for the user to receive the intervention. The suggestions could be tailored to the participant's context within 90 seconds of a decision point, and messages could be delivered consistently and at appropriate times. Further, anticipating the possibility of a lost internet connection at the next decision point, our system collected contextual data and ``pre-fetched'' an activity suggestion from the server 30 minutes prior to randomization. This suggestion would be delivered at the decision point if the participant was randomized to receive treatment and did not have an internet connection. This maximized the chances that activity suggestions could be successfully randomized and delivered even if the participant lost connectivity. 

Phone-side randomization also has downsides. In particular, it requires extensive safeguards, such as ``handshakes'' which check for a stable connection between phone and server, to ensure the integrity of transmitted data. As an example, the order of the questions in the HeartSteps evening survey was randomized by the phone. The app was designed to send participants' responses to the server immediately after each question was answered. Unfortunately, in several cases, technical difficulties led to incomplete survey responses being recorded which prevented analysts from knowing the precise order of questions that should have been given to the participant. Furthermore, the stored results of randomization for the planning intervention were found to be unreliable and were frequently missing due to data loss. Server-side randomization would not have this issue.

Making decisions on the server can facilitate the automatic creation of clean (and possibly more complete) datasets, but introduces some ambiguity in the timing and tailoring of the intervention. Data tables on the server can be pre-built: the system can be designed to expect one row per decision per participant, and randomization status is always known. These pre-built tables can be filled in over the course of the study. This eliminates possible duplication due to time zone changes or other bugs (see Section~\ref{sec:time}). However, push notifications delivered by Apple (iOS) or Google (Android) from the cloud can be delayed, resulting in uncertainty in the time at which the intervention was delivered to the participant. This delay could also cause the intervention to be tailored to a context which has changed and is no longer appropriate.

\section{Unavailability and Missing Data}
\label{sec:unavail-missingness}
A notable feature of an MRT is the incorporation of ``availability'', which protocolizes the notion that it may only be appropriate to provide treatment when the participant is in certain contexts or ``states"~\citep{Klasnja2015}. Participants may be considered ``unavailable'' for treatment for reasons of safety, burden/annoyance, or feasibility. For example, in HeartSteps, participants were considered unavailable if they were currently driving, did not have an active internet connection, they manually turned off the intervention, or were walking within 90 seconds of a decision point, as measured by their phone~\citep{Smith2017}.

By design, if a participant is unavailable at a decision point then the the participant is not randomized at that time~\citep{Boruvka2017,Klasnja2015,Liao2016}. However, it may be of interest to investigate reasons for unavailability. In HeartSteps, for example, we hypothesized that the proportion of suggestion decision points at which participants were unavailable due to walking would increase over time. This would suggest the intervention helped them become more active without the need for prompts. Investigating this, however, requires collecting the same data from both unavailable and available participants, even though treatment cannot be delivered for the former. 

Ideally, when the participant is unavailable at a certain decision point, the data collection system will be able to identify and record the specific reason for that unavailability. This is partially facilitated by a clear, protocolized definition of availability, which is likely to be multifaceted and time-sensitive~\citep{Smith2017}, and partially by collecting the usual set of features at unavailable decision points. At each decision point, then, an ideal system would make a determination as to participant availability, output an availability indicator for that participant and, for unavailable participants, indicate which availability criterion/criteria were not met.

Though unavailability and missingness are distinct concepts, ideal MRT data collection will handle them in similar ways. As much as possible, systems should be designed to capture data necessary to pinpoint reason(s) for data missingness. This is critical for making valid inference~\citep{Rubin1976}, and identifying missingness mechanisms can aid analysts in discovering technical issues which may impact analyses or in identifying participant disengagement. For example, if participants do not respond to intervention components which invite or require an action (e.g., daily planning in HeartSteps), the system should always store ``no response'' rather than a blank entry. This can tell the analyst that the intervention was delivered, but the participant did not engage with it, thus allowing the analyst to distinguish between missing data due to disengagement and missing data due to, e.g., technical issues that prevented intervention delivery. 

Some reasons for unavailability or missing data may be difficult or impossible to ascertain with certainty. For example, to preserve battery life, fitness trackers may only record positive step counts; that is, if a participant took zero steps in a certain period, the tracker will not report anything for that time. In some cases, then, it may not be possible to tell if gaps in step counts are due to true inactivity, or if the participant stopped wearing the tracker. Because of this, primary analyses of HeartSteps data used a zero-imputed version of Jawbone step count~\citep{Klasnja2018}. Having redundant data streams can sometimes help in such cases---in HeartSteps, Google Fit provided another source of step count data, allowing for sensitivity analyses in which missing Jawbone step counts were singly imputed from Google Fit---but resolving all cases of uncertain missingness may still prove difficult. 
To anticipate and mitigate the various scenarios that can result in missing data before the study starts, scientists should employ systematic software testing and deployments of beta versions of the intervention.

\section{Mitigating Inadvertent Participant Errors}
\label{sec:human}
Passive data collection through mobile devices like smartphones can be idiosyncratic in ways that would not arise in more traditional clinical settings. Primarily, it is possible for participants to (inadvertently) fail to provide data, potentially for long periods of time. This is a serious issue in MRTs because of the density of the data collected. Study designers should develop protocols to identify when participants are not contributing data, and take appropriate steps to correct this.

There are several ways in which a participant's data collection could be interrupted. Phones might lose power or be manually turned off, thus limiting the ability for sensors to collect data. Generally, apps are not notified when the phone is shutting down, and so this situation may be impossible to identify. The operating system may also shut down any background processes of the MRT app to free up memory or improve battery life, stopping data collection until the user manually opens the application again. Batteries in wearable sensors may die, requiring the participant to notice their sensor is dead, charge it, and put it back on. All of these situations create missingness for which the cause is difficult to identify from the data. Similarly, a participant might turn off Bluetooth, which is often required to sync data from sensors to the phone, or GPS, necessary for location detection. In some cases, data might be stored on the sensor but not synced to the server, so step counts, for example, might need to be recovered after a Bluetooth connection is restored. Location data, however, is likely not recoverable if GPS is turned off. 

Missingness due to data loss should be mitigated as much as possible. App developers should attempt to anticipate situations that could lead to data loss, such as attempting to deliver the intervention over a WiFi network with a ``captive portal'' that requires users to accept terms of service before connecting (e.g., hotel WiFi). With HeartSteps, we discovered that some participants were used to closing apps by ``swiping them away'' from the phone's multitasking menu. If the participant closed the HeartSteps app this way before it ensured that data sent from the app was received by the server, data loss could result. Both of these situations might be mitigated by phone-to-server ``handshakes'', which track data exchange between the phone and the server and can signal to the phone that the exchange was successfully completed so that the phone-side data can be discarded.

\section{Telling Time in MRT Data}
\label{sec:time}
MRTs require careful consideration of time, especially as it relates to the treatments delivered and participants' experiences with those treatments. Special attention must be given to collecting time stamps, particularly for events related to treatment delivery. Not only is this crucial for aligning data from multiple sources (e.g., step count data from the fitness tracker and information from the phone), but it also allows for easier troubleshooting in case of bugs. Careful collection of time stamps allows for the reconstruction of the participant's experience with the MRT.

The ability to infer the participant's ``timeline'' from data is critical. Since data collection is automated and no code is perfect, technical difficulties \textit{will} arise. Time stamps can identify and troubleshoot duplicated records, and can allow for the linking of data from different components of the intervention. Immediately translating all timestamps into Coordinated Universal Time (UTC) helps avoid problems with participants changing time zones and Daylight Savings Time. UTC time stamps allow for the cleanest possible reconciliation of different variables that need to be connected to each other by time stamps.

In HeartSteps, information about message delivery and the participant's response to the message were kept in separate tables on the central server, and later linked by participant ID and time stamps. Careful collection of time stamps can identify delays in the intervention delivery system: in MRTs that use server-side randomization, the time of the decision to provide treatment will likely not be the time at which the treatment is delivered via the participant's phone. Recording the time at which this delivery does occur, or some surrogate thereof, can indicate the proper time at which to start measuring the proximal outcome.

For longer studies, participant travel during the study should also be considered. One of the advantages of mobile intervention delivery is that it can follow participants as they, for example, go on vacation. However, due to the highly time-sensitive nature of the data collected in an MRT, changes in time zones can result in aberrant behavior in the app and/or data collection systems, such as repeated or missed decision points. As such, time stamps should be stored with the participant's current time zone when possible (note that this might require access to participants' locations). Smartphone system times may not adjust to time zone changes unless the phone is rebooted. If this is the case, duplication of decision points may not be an issue, but if the decision times are chosen in advance by the user, improperly-timed treatments may be jarring. For example, consider a participant traveling from the east coast of the US to Hawaii---a six hour time difference. If the interventions are delivered according to a system time which is not sensitive to time zone, the participant might receive a suggestion tailored for the evening around noon, or for very different weather conditions.

Study designers should decide \textit{a priori} how to handle participant travel. One approach might be to exclude time spent traveling from an individual's data, though this requires strong scientific justification and a comprehensive definition of what constitutes ``travel''. Any information needed to assess whether the user is traveling should also be recorded. Alternatively, if the app and intervention can handle time zone changes without creating jarring experiences like duplicated or mistimed decision points, time could be conceptualized as the participant's local time and decision points indexed according to this local time.

Just as important as proper time stamp collection is consideration of the various time scales on which variables are measured, which are likely reflective of how dynamic or changeable these features are over time. Some features might be collected only at baseline (e.g., age) or monthly (e.g., self-efficacy); others, daily (e.g., how typical a day the participant had) or even at every decision point (e.g., weather). In preparing data for analysis, these time scales should be carefully accounted for.  For example, in HeartSteps, weather conditions were collected at every decision point. In determining whether weather moderates the effect of the daily planning intervention on daily step count, average temperature or total precipitation throughout the day might be used as a summary measure. On the other hand, when merging a daily-level variable (e.g., self-reported stress level) onto finer data (such as activity suggestions), care should be taken so that all decision points on the same day are associated with the same, most-recent daily observation. Recall that adjustment for post-treatment variables can lead to biased causal inference~\citep{Rosenbaum1984}.

\section{Discussion}
\label{sec:discussion}
Mobile health interventions delivered through smartphones have the potential to effect meaningful change in individuals' lives. The power and prevalance of mobile devices has led to a rapid pace in innovation and a growing interest in harnessing them to improve health. The micro-randomized trial allows for the collection of data which can be used to construct optimized just-in-time adaptive interventions and make causal inference about the effects of intervention components on a proximal outcome. However, care must be taken to ensure that this data is high-quality. 

To this end, we have presented a series of recommendations in several key areas which, if implemented, can help scientists collect and manage data which can be used to assess the effectiveness of momentary, mobile interventions. The central theme of these recommendations is that careful planning is required before the trial begins to ensure that the proper data is collected in a robust way. Passive data collection systems in MRTs are not able to document their procedures (as a human might) other than through their source code, so testing the application is critical.

Designing a smartphone app and data collection system to address scientific questions involves different needs than does typical app development. Using data from smartphones or sensors in an MRT to perform causal inference requires that careful attention is paid to precisely how the data is gathered. The checklist presented in the Appendix may be a useful start to ensure that appropriate, high-quality data can be collected in a general MRT. The recommendations presented here can allow scientists to focus more on their questions of interest, and spend less time working with disorderly data.

\section*{Acknowledgements}
Research reported in this publication was supported by NIAAA, NIDA, NIBIB, and NHLBI of the National Institutes of Health under award numbers R01AA023187, P50DA039838, U54EB020404, and R01HL125440. The content is solely the responsibility of the authors and does not necessarily represent the official views of the National Institutes of Health. The authors wish to thank Dr. Audrey Boruvka for her work in managing data from HeartSteps, as well as the two anonymous reviewers for their thoughtful comments.

%\printbibliography

\newpage
\section*{Appendix: Checklist for Preparing a Data Collection System for an MRT}
Here, we summarize the ideas presented above into a checklist which can be used when collaborating with app developers to design a data collection system for an MRT. The following is a general guide to key steps which should be addressed in the planning stages of most MRTs to improve the quality of data collected.

\begin{enumerate}
	\item \textit{Proximal Outcome.} Know how the proximal outcome is measured and collected, and on what time scale(s) it is available. (Section~\ref{sec:proximal-outcome})
	
	\item \textit{Randomization Agent.} Decide whether randomization should occur on the participant's phone or on a central server, being careful to weigh advantages and disadvantages of both in the context of the MRT being designed. (Section~\ref{sec:agents})
	
	\item \textit{Treatment Delivery.} Ensure robust collection of data on treatment delivery at each decision point. This includes the results of each randomization and the probability with which individuals were randomized to receive treatment. (Section~\ref{sec:txt-delivery})
	
	\item \textit{Contextual Data.} List contextual variables which may be of interest for tailoring the intervention, or as possible moderators for later analysis. Ensure that all contextual variables are collected regardless of the decision to treat or not treat. Assess the time scales on which these features are available, and determine how to ensure those used for tailoring are kept ``fresh''.
	Be sure to consider participant privacy when choosing what to collect and storing the data. (Sections~\ref{sec:context} and \ref{sec:privacy})
	
	\item \textit{Time Stamps.} As much as possible, collect the times at which key events in the trial occurred, such as randomization times, and times at which interventions were delivered to and/or seen by the participant. Immediately translate all time stamps into UTC. If location is available, collecting the participant's local time zone with every time stamp can help troubleshoot issues caused by travel. (Section~\ref{sec:time})
	
	\item \textit{Unavailability and Missing Data.} Develop an appropriately comprehensive definition of unavailability that allows the study app to accurately record \textit{why} participants were unavailable. Ensure that all contextual data is collected at each decision point regardless of availability status. Anticipate and address sources of missing data by understanding how sensors record a null event (e.g., zero steps), and by storing ``no response'' when an intervention is delivered but the participant does not engage. Use phone-to-server ``handshakes'' to ensure data is transfered successfully before it is discarded on the phone. As much as possible, missingness mechanisms should be identified and recorded in the data. (Section~\ref{sec:unavail-missingness})
	
	\item \textit{Understanding Participants.} List and design around common issues participants might have with the study app and/or phone. Protocolize strategies for data recovery when, say, a participant turns Bluetooth off or the battery on his/her sensor dies. Carefully pilot test the app before launching the study to help identify bugs which may arise in day-to-day usage. (Section~\ref{sec:human})
\end{enumerate}

\end{document}